\documentclass[a4paper]{article}

\usepackage{INTERSPEECH2021}
\usepackage{amsmath,bm,cite} %
\usepackage{makecell}
\usepackage{arydshln}
\usepackage{etoolbox,siunitx}
\robustify\bfseries
\usepackage{balance}

\usepackage{pifont}

\makeatletter
\def\adl@drawiv#1#2#3{%
        \hskip.5\tabcolsep
        \xleaders#3{#2.5\@tempdimb #1{1}#2.5\@tempdimb}%
                #2\z@ plus1fil minus1fil\relax
        \hskip.5\tabcolsep}
        
\newcommand{\cdashlinelr}[1]{%
  \noalign{\vskip\aboverulesep
           \global\let\@dashdrawstore\adl@draw
           \global\let\adl@draw\adl@drawiv}
  \cdashline{#1}
  \noalign{\global\let\adl@draw\@dashdrawstore
           \vskip\belowrulesep}}
           
\newcommand{\myhdashline}{%
  \noalign{\vskip\aboverulesep
           \global\let\@dashdrawstore\adl@draw
           \global\let\adl@draw\adl@drawiv}
  \hdashline
  \noalign{\global\let\adl@draw\@dashdrawstore
           \vskip\belowrulesep}}
\makeatother

\title{Dual Causal/Non-Causal Self-Attention\\for Streaming End-to-End Speech Recognition}
\name{Niko Moritz, Takaaki Hori, Jonathan Le Roux}

\address{Mitsubishi Electric Research Laboratories (MERL), Cambridge, MA, USA}
\email{\{moritz, thori, leroux\}@merl.com}

\begin{document}

\setlength{\abovedisplayskip}{4pt}
\setlength{\belowdisplayskip}{4pt}
\setlength{\abovecaptionskip}{3pt}
\setlength{\belowcaptionskip}{-3pt}
\setlength{\textfloatsep}{5pt}
\setlength{\parindent}{1em}

\maketitle
\begin{abstract}
  Attention-based end-to-end automatic speech recognition (ASR) systems have recently demonstrated state-of-the-art results for numerous tasks. However, the application of self-attention and attention-based encoder-decoder models remains challenging for streaming ASR, where each word must be recognized shortly after it was spoken. In this work, we present the dual causal/non-causal self-attention (DCN) architecture, %
  which in contrast to restricted self-attention prevents the overall context to grow beyond the look-ahead of a single layer when used in a deep architecture.
  DCN is compared to chunk-based and restricted self-attention using streaming transformer and conformer architectures, showing improved ASR performance over restricted self-attention and competitive ASR results compared to chunk-based self-attention, while providing the advantage of frame-synchronous processing. Combined with triggered attention, the proposed streaming end-to-end ASR systems obtained state-of-the-art results on the LibriSpeech, HKUST, and Switchboard ASR tasks. 
\end{abstract}

\section{Introduction}

In recent years, end-to-end (E2E) automatic speech recognition (ASR) has superseded conventional hybrid DNN/HMM solutions \cite{LiZML2020,SainathHLN2020}. Three main reasons can be noted for this development: 1) E2E ASR solutions have a much less complex training pipeline, 2) the inference process of E2E ASR systems can be optimized more easily, and 3) E2E ASR models have proven to achieve competitive or better performance compared to conventional hybrid solutions.
The most popular E2E ASR models are based on the connectionist temporal classification (CTC) \cite{Graves2006}, the RNN transducer (RNN-T) \cite{Graves12}, and/or an attention-based encoder-decoder architecture \cite{Chorowski2015,Watanabe2017}.
Besides the progress that has been made in building E2E ASR systems %
\cite{ChiuSWP2018,PrabhavalkarRSL17}, more advanced neural network architectures are an important factor in the dramatic improvement of ASR results in the past decade.
The most successful ASR architectures are transformer and conformer-based \cite{GulatiCQY2020,GuoBCC2020recent}, which both apply self-attention and source (or encoder-decoder) attention for analyzing temporal information of a sound signal.
However, both attention types typically require a full input sequence corresponding to an entire speech utterance, e.g., extracted using speech activity detection, even for recognizing an ASR output corresponding to the beginning of a long utterance.

Several methods have been proposed to improve the streaming properties of source attention in order to generate ASR outputs with a controlled delay. The neural transducer \cite{JaitlyLVSSB16,SainathCPKWNC17} and blockwise synchronous beam search \cite{TsunooKW21} both apply the source attention on chunks of encoder frames with a fixed striding instead of on the full encoder sequence corresponding to a complete utterance. 
Other methods perform the chunking with an adaptive stride, e.g., by computing a selection probability such as Monotonic Chunkwise Attention (MoChA) \cite{ChiuR18,AriCMC2019} or by detecting word boundaries using a scout network \cite{WangWLL2020}. In \cite{MoritzHR19}, we introduced triggered attention (TA), which exploits the temporal alignment properties of CTC to identify encoder frames containing relevant information for an ASR output and to trigger an attention-based ASR decoder at such time positions. Unlike other solutions, TA provides a frame-synchronous one-pass decoding algorithm with joint CTC/attention scoring \cite{MoritzHR19c,MoritzHLR20}. 

For streaming E2E ASR, restricted self-attention (RSA) and chunk-based self-attention (CSA) are often used \cite{zhang2020transformer,MoritzHLR20,DongWX2019,MiaoCGZ2020}, where the latter has been shown to achieve better ASR performance but with the drawback of working in a non-frame-synchronous manner and requiring higher computational costs, which is often addressed by compressing the past context using context inheritance or augmented memory \cite{WuWSY2020,TsunooKW21}. 

In this work, we present the dual causal/non-causal self-attention (DCN) architecture for streaming E2E ASR. DCN provides frame-synchronous processing with a fixed look-ahead size without causing an increasing receptive field for a growing number of consecutive DCN layers, which is unlike the delay aggregation of deep RSA architectures \cite{MoritzHLR20,zhang2020transformer}. %
We combine DCN-based encoder models with TA for streaming E2E ASR using transformer- and conformer-based neural network architectures.
RSA, DCN, and CSA-based ASR results are compared using the LibriSpeech, HKUST, and Switchboard tasks. For all ASR tasks, the proposed streaming E2E ASR system achieves state-of-the-art results.
In addition, the decoding emission delays of various TA-based models are analyzed, demonstrating that TA decoding is well aligned with the true time positions of words and that streaming models with less look-ahead are more prone to delaying ASR decoding outputs.

\vspace{-0.1cm}
\section{System Architecture}
\vspace{-0.1cm}

ASR systems in this work make extensive use of the scaled dot-product attention mechanism, which can be written as
\begin{equation}
    \mathrm{Attention}(Q, K, V) = \mathrm{Softmax}\left(\dfrac{Q K^\mathsf{T}}{\sqrt{d_k}}\right)V, \label{eq:mhatt} 
\end{equation}
where $Q \in \mathbb{R}^{n_q \times d_q}$, $K \in \mathbb{R}^{n_k \times d_k}$, and $V \in \mathbb{R}^{n_v \times d_v}$ are the queries, keys, and values, where the $d_*$ denote dimensions and the $n_*$ denote sequence lengths, $d_q=d_k$, and $n_k=n_v$. Multiple attention heads may be used by
\begin{align}
    \mathrm{MHA}(\hat Q, \hat K, \hat V) &= \mathrm{Concat}(\mathrm{Head}_1, \dots, \mathrm{Head}_{d_h}) W^\text{H}, \label{eq:mha}\\
    \text{and } \mathrm{Head}_i &= \mathrm{Attention}(\hat Q W_i^Q, \hat K W_i^K, \hat V W_i^V),  \label{eq:att_heads}
\end{align}
where $\hat Q$, $\hat K$, and $\hat V$ are inputs to the multi-head attention (MHA) layer, $\mathrm{Head}_i$ represents the output of the $i$-th attention head for a total number of $d_h$ heads, $W_i^Q \in \mathbb{R}^{d_\mathrm{model} \times d_q}$, $W_i^K \in \mathbb{R}^{d_\mathrm{model} \times d_k}$, $W_i^V \in \mathbb{R}^{d_\mathrm{model} \times d_v}$ as well as $W^H \in \mathbb{R}^{d_hd_v \times d_\mathrm{model}}$ are trainable weight matrices with typically $d_k=d_v=d_\mathrm{model}/d_h$, and $\mathrm{Concat}$ denotes concatenation.

In this work, transformer and conformer architectures are employed in a joint CTC/attention based E2E ASR system \cite{KaritaYWD19,GuoBCC2020recent}. %
Both architectures first extract 80-dimensional log-mel spectral energies plus 3 extra features for pitch information \cite{watanabe2018espnet}. The derived feature sequence $X$ is processed by a two-layer convolutional neural network (CNN) module, which generates feature sequence $X_0$ with a frame rate of 40~ms \cite{MoritzHLR20}. %

For $e=1,\dots,E$ blocks, the \textbf{transformer} encoder applies multi-head self-attention ($\mathrm{MHA}$), a feed-forward neural network module ($\mathrm{FF}$), and layer normalization ($\mathrm{LN}$) as follows:
\begin{align}
    \tilde{X}_e &= \mathrm{LN}_{e,1}(X_{e-1}) \label{eq:sa_LN}\\
    \overline{X}_e &= X_{e-1} + \mathrm{MHA}_{e}(\tilde{X}_{e}, \tilde{X}_{e}, \tilde{X}_{e}), \label{eq:sa_transf} \\
    X_{e} &= \overline{X}_e + \mathrm{FF}_e(\mathrm{LN}_{e,2}(\overline{X}_e)), 
\end{align}
where each $\mathrm{FF}_e$ module consists of two linear layers of inner dimension $d_\mathrm{in}$ with a ReLU non-linearity in between. A sinusoidal positional encoding (PE)  \cite{VaswaniSPUJGKP17} is added to $X_0$ prior to feeding it to the transformer neural network. 
Dropout with a probability of $0.1$ is used after self-attention and after each layer of the $\mathrm{FF}$ module. %

The \textbf{conformer} encoder is composed of $e=1,\dots,E$ conformer blocks that can be written as
\begin{align}
    \tilde{X}_{e} &= X_{e-1} + \frac{1}{2}\mathrm{FF}_{e,1}(\mathrm{LN}_{e,1}(X_{e-1})), \\
    \tilde{\tilde{X}}_e &= \mathrm{LN}_{e,2}(\tilde{X}_{e}) \\
    \overline{X}_e &= \tilde{X}_{e} + \mathrm{MHA}^\mathrm{pos}_{e}(\tilde{\tilde{X}}_{e}, \tilde{\tilde{X}}_{e}, \tilde{\tilde{X}}_{e}), \label{eq:sa_conf} \\
    \overline{\overline{X}}_e &= \overline{X}_e + \mathrm{Conv}_{e}(\mathrm{LN}_{e,3}(\overline{X}_e)), \label{eq:conv_module} \\
    X_{e} &= \overline{\overline{X}}_e + \frac{1}{2}\mathrm{FF}_{e,2}(\mathrm{LN}_{e,4}(\overline{\overline{X}}_e)), 
\end{align}
where $\mathrm{MHA}^\mathrm{pos}$ denotes multi-head self-attention with relative positional encoding \cite{GuoBCC2020recent}, $\mathrm{Conv}$ is a convolution module, and both $\mathrm{FF}$ module architectures are similar to $\mathrm{FF}$ of the transformer. Dropout with probability $0.1$ is used after $\mathrm{MHA}^\mathrm{pos}$, $\mathrm{Conv}$, and each layer of $\mathrm{FF}$. %
The $\mathrm{Conv}$ module starts with a pointwise convolution layer and a gated linear unit (GLU) followed by a 1-D depthwise convolution, batch normalization, Swish activation, and another pointwise convolution layer \cite{GuoBCC2020recent}.

{\allowdisplaybreaks
A CTC and an attention decoder branch are jointly trained using the multi-objective loss function
\begin{equation}
\mathcal{L} = -\gamma \log p_\text{ctc} - (1-\gamma) \log p_\text{dec}, \label{eq:loss}
\end{equation}
where $p_\text{ctc}$ and $p_\text{dec}$ denote the CTC and attention decoder loss with hyperparameter $\gamma$ controlling the weighting between the two.
In the \textbf{CTC} branch, the encoder output $X_E$ is layer normalized and frames are projected to a probability distribution of the size of the ASR labels plus one for the CTC blank symbol by using a linear layer followed by a softmax function.
In the \textbf{decoder} branch, a decoder model is trained using
\begin{equation}
  p_{\text{dec}}(Y|X_E) = \prod_{l=1}^{L} p(y_l | \bm y_{1:l-1}, X_E) \label{eq:dec_objf}
\end{equation}
with label sequence $Y=(y_1,\dots,y_L)$, label subsequence $\bm y_{1:l-1}=(y_1,\dots,y_{l-1})$, and the encoder output sequence $X_E$. The posterior probability $p(y_l | \bm y_{1:l-1}, X_E)$ is computed by the decoder model as follows:
\begin{align}
  \bm z_{1:l}^{0} &= \textsc{Embed}(\langle \text{s} \rangle, y_{1}, \dots, y_{l-1}) + \mathrm{PE}, \label{eq:dec_first} \\
  \tilde{X}_{E} &= \mathrm{LN}(X_{E}), \\
  \hat{\bm z}_{1:l}^{d} &= \mathrm{LN}_{d,1}(\bm z_{1:l}^{d-1}), \\
  \overline{\bm z}^d_{l} &= \bm z_{l}^{d-1} + \mathrm{MHA}_d^\mathrm{self}(\hat{\bm z}_{l}^{d},\hat{\bm z}_{1:l}^{d},\hat{\bm z}_{1:l}^{d}), \\
  \overline{\overline{\bm z}}^d_{l} &= \overline{\bm z}^d_{l} + \mathrm{MHA}_d^\mathrm{src}(\text{LN}_{d,2}(\overline{\bm z}^d_{l}), \tilde{X}_E, \tilde{X}_E)), \label{eq:dec_att} \\
  \bm z_{l}^{d} &= \overline{\overline{\bm z}}^d_{l} + \mathrm{FF}_d(\mathrm{LN}_{d,3}(\overline{\overline{\bm z}}^d_{l})),
  \label{eq:dec_last}
\end{align}
for $d=1,\dots,D$, where $D$ denotes the number of decoder blocks, $\mathrm{PE}$ is a sinusoidal positional encoding, \textsc{Embed} converts the input label sequence $(\langle \text{s} \rangle, y_{1}, \dots, y_{l-1})$ into a sequence of trainable embedding vectors $\bm z_{1:l}^{0}$, and $\langle \text{s} \rangle$ denotes a start of sentence symbol.
Finally, output vector $\bm z_{l}^{D}$ is further processed using LN, a linear layer, and a softmax function to compute a probability distribution over the ASR labels.
}

For \textbf{streaming} E2E ASR, model components that process temporal information must be restricted to only use a limited future context for reducing the output latency.
In the presented architectures, the self-attention layers of Eqs.~(\ref{eq:sa_transf}) and (\ref{eq:sa_conf}), the $\mathrm{Conv}$ module of Eq.~(\ref{eq:conv_module}), and the source attention of Eq.~(\ref{eq:dec_att}) typically require a wide temporal context. RSA  \cite{VaswaniSPUJGKP17,MoritzHLR20,zhang2020transformer}, CSA \cite{DongWX2019,TsunooKW21}, and the dual causal/non-causal (DCN) self-attention introduced below are used in this work to limit the algorithmic latency of self-attention. For streaming ASR with the conformer architecture, a causal $\mathrm{Conv}$ module is used by restricting the convolution window of the 1-D depthwise convolution to 16 past frames plus the current frame. Finally, streaming ASR with source attention is realized using TA.

\section{Dual Causal/Non-Causal Self-Attention}

\begin{figure}[t]
  \centering
  \centerline{\includegraphics[width=\linewidth]{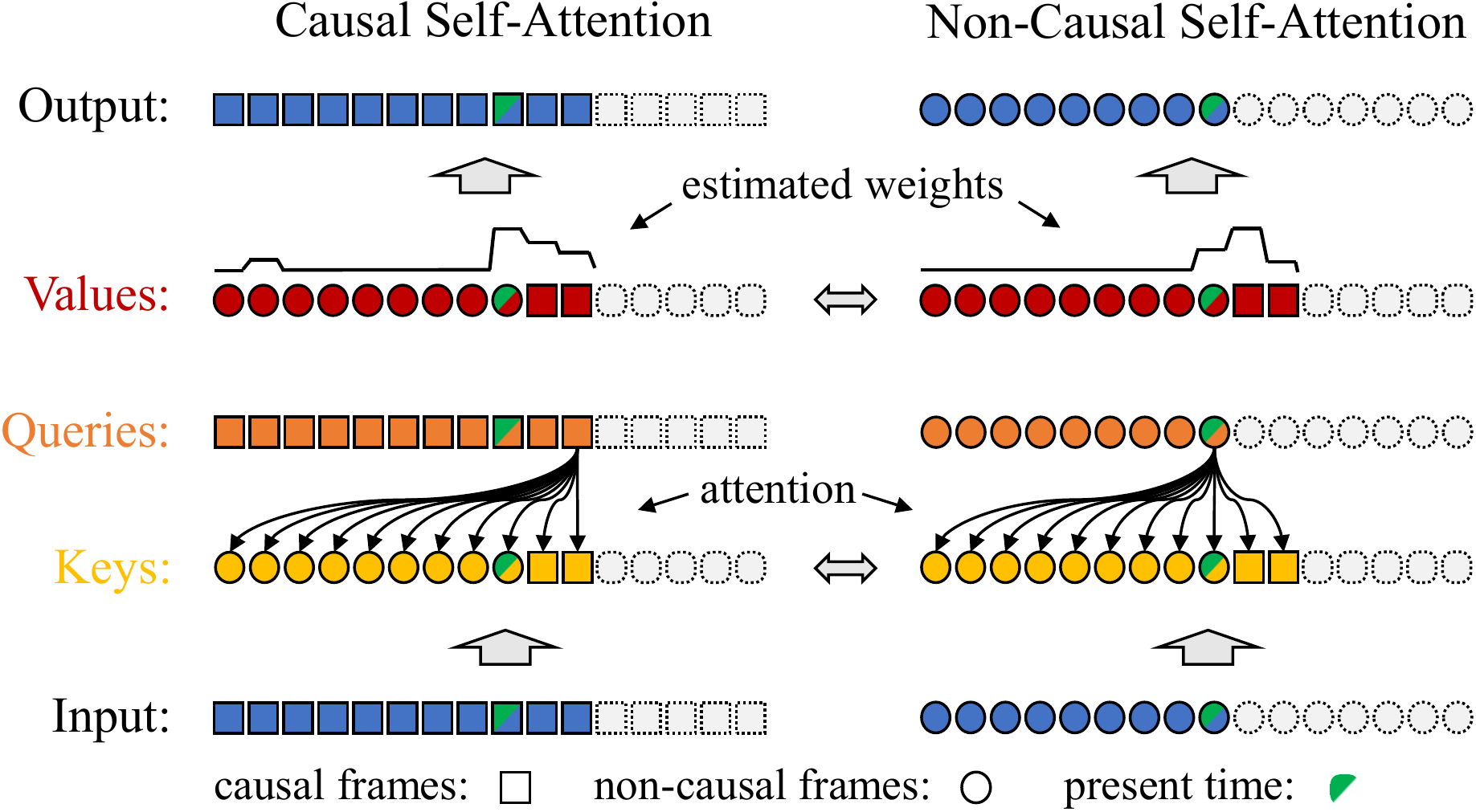}} %
  \caption{Dual causal/non-causal self-attention (DCN) with 2 look-ahead frames.} \vspace{.1cm}
\label{fig:dual-self-att}
\end{figure}

The most widely used streaming self-attention architectures are restricted self-attention (RSA) and chunk-based self-attention (CSA). RSA uses a limited number of look-ahead frames per self-attention layer \cite{MoritzHLR20}, which results in a growing receptive field when a series of RSA layers are used.
The reason for the growing delay is that frames input to RSA may be non-causal, i.e., they were computed requiring future context, and thus delays add up for each additional layer that uses a look-ahead for computing an output.
To avoid this build up, DCN processes two sequences of causal and non-causal frames in parallel as shown in Fig.~\ref{fig:dual-self-att}.
For the first encoder block, the input feature sequence $X_0$ is simply duplicated to derive a causal sequence $X^\mathrm{c}_0$ and a non-causal sequence $X^\mathrm{nc}_0$ of frames.
Next, both sequences are processed by two parallel RSA processes, one with causal RSA using zero look-ahead frames and one with non-causal RSA using a fixed number of look-ahead frames.
DCN first transforms the causal and non-causal input frames into causal and non-causal key, value, and query frames.
Key and value frames are interchanged between the parallel causal and non-causal RSA processes such that for the non-causal RSA process the look-ahead frames correspond to causal frames, and for the causal RSA process the frames that are further into the past than the look-ahead size of the non-causal frames are non-causal frames. This ensures that both RSA processes do not attend to any frames that use information beyond the attention context.
Since two sequences must be forwarded to the next DCN layer, both sequences must be processed by all modules of an encoder block, which increases the computational costs. However, the computational cost is approximately equivalent to CSA with 50\% chunk overlap, which also almost doubles the number of frames that are being processed.

Finally, at the output of the encoder, the non-causal encoder sequence $X^\mathrm{nc}_E$ is forwarded to the CTC and decoder branches, while the causal sequence $X^\mathrm{c}_E$ is dropped. In order to enforce a consistency between the causal and non-causal sequences, we explored inplace knowledge distillation (KD) \cite{yu2019universally,yu2021dualmode} between $X^\mathrm{c}_E$ (student) and $X^\mathrm{nc}_E$ (teacher) using the mean squared error (MSE) loss. In our experiments, the MSE loss is multiplied with a weight of $1.0$ and added to Eq.~(\ref{eq:loss}), unless otherwise noted in the results section below.
All model parameters are shared for processing the causal and non-causal frames in the encoder, except parameters of the normalization layers.

\section{Triggered Attention}

For streaming E2E ASR, we use the triggered attention (TA) technique, which exploits the alignment properties of CTC to enable frame-synchronous decoding of encoder-decoder based ASR models \cite{MoritzHR19,MoritzHR19c}. TA training uses CTC-based forced alignment to determine the temporal position of the label $y_l$ for $l\!=\!1,\dots,L$ in the encoder output sequence $X_{E}\!=\!(\bm x^E_1,\dots,\bm x^E_N)$ in order to condition the source attention mechanism of the decoder on previous encoder frames plus a fixed number of look-ahead frames $\varepsilon^\text{dec}$ relative to the determined label positions $n'_l$. The TA loss can be written as
\begin{equation}
  p_{\text{ta}}(Y|X_E) = \prod_{l=1}^{L} p(y_l | \bm y_{1:l-1}, \bm x_{1:\nu_l}^E) \label{eq:trig_att}
\end{equation}
with $\nu_l = n'_l + \varepsilon^\text{dec}$, where $n'_l$ denotes the position of the first occurrence of label $y_l$ in the CTC forced alignment sequence \cite{MoritzHR19,MoritzHR19c}, and $\bm x_{1:\nu_l}^E=(\bm{x}_1^E,\dots,\bm{x}_{\nu_l}^E)$ corresponds to the truncated encoder sequence. %
The term $p(y_l | \bm y_{1:l-1}, \bm x_{1:\nu_l}^E)$ represents the posterior probability function of the TA decoder that substitutes Eq.~(\ref{eq:dec_objf}) for streaming ASR and which is computed similarly to Eqs.~(\ref{eq:dec_first}) to (\ref{eq:dec_last}) but with the restricted encoder sequence $\bm x_{1:\nu_l}^E$ deployed in Eq.~(\ref{eq:dec_att}).

At inference time, we use the TA decoding algorithm of \cite{MoritzHR19c,MoritzHLR20}, which extends the frame-synchronous CTC prefix beam search algorithm with TA-based on-the-fly rescoring. 

\section{Experiments}

\subsection{Settings}

ASR tasks used for the present experiments are the LibriSpeech corpus of read English audio books \cite{librispeech}, the Switchboard (SWBD) corpus of conversational English telephone speech \cite{swbd}, and the Hong Kong University of Science and Technology (HKUST) corpus of Mandarin telephone speech \cite{hkust}. %

Hyperparameters of the transformer and conformer models are set to $d_\mathrm{model}=256$, $d_\mathrm{ff}=2048$, $d_h=4$, $E=12$, and $D=6$ for HKUST and SWBD, while $d_\mathrm{model}$ and $d_h$ are increased to $512$ and $8$ for LibriSpeech.
A symmetric depthwise convolution window of size 31 frames is used by the $\mathrm{Conv}$ module of the full-sequence conformer encoder. For streaming ASR, a causal (left-sided) convolution window of size 17 is used instead.
The Adam optimizer with $\beta_1=0.9$, $\beta_2=0.98$, $\epsilon=10^{-9}$, and learning rate scheduling similar to \cite{VaswaniSPUJGKP17} with 25000 warmup steps is applied for training.
The learning rate factor is set to $5.0$ and the number of training epochs amounts to $50$ for HKUST and to $100$ for LibriSpeech as well as SWBD. 
Weight factor $\gamma$, which is used to balance the CTC and decoder objectives, is set to $0.3$ for LibriSpeech and HKUST and to $0.2$ for SWBD. Label smoothing with a penalty of $0.1$ and SpecAugment are used for all experiments \cite{park2019specaugment}.
A task specific RNN-based language model (LM) is trained via stochastic gradient descent using the official training text data of each task \cite{librispeech,hkust,swbd} and employed via shallow fusion during decoding. For HKUST and SWBD, the RNN-LM consists of 2 LSTM layers with 650 units each. For LibriSpeech, 4 LSTM layers with 2048 units each are used. When indicated in the LibriSpeech results, a Transformer-based LM (Tr-LM) with 16 layers is used instead. ASR output labels consist of a blank token plus 5,000 (LibriSpeech) or 2,000 (SWBD) subwords obtained by the SentencePiece method \cite{KudoR18}. 3,653 character-based output symbols are use for the Mandarin HKUST task.

In this work, TA-based models are obtained by fine-tuning a pre-trained model with a full-sequence based decoder for 10 epochs using the Adam optimizer without learning rate scheduling. \textit{Fine-tuning} has proved to be very effective in this work, reducing the training time and improving ASR accuracy compared to TA models trained from scratch such as in \cite{MoritzHLR20}.

For offline decoding, the LM weight/CTC weight/beam size are set to 0.6/0.4/20 (LibriSpeech), 0.3/0.5/10 (HKUST), and 0.3/0.3/20 (SWBD). For frame-synchronous TA decoding \cite{MoritzHR19c,MoritzHLR20}, the CTC LM weight $\alpha_0$/CTC weight $\lambda$/LM weight $\alpha$/pruning width $\theta_1$/pruning width $\theta_2$/insertion bonus $\beta$/pruning size $K$/beam size $P$ are set to 0.8/0.4/0.6/16.0/6.0/2.0/300/30 (LibriSpeech), 0.4/0.6/0.2/8.0/4.0/1.0/200/20 (HKUST), and 0.4/0.6/0.3/8.0/2.0/1.0/200/20 (SWBD).

\subsection{Chunk-based self-attention}

The CSA mechanism first segments the whole utterance into chunks using a hop size of 50\%. In order to enable a fair comparison to RSA and DCN, CSA in this work attends to all frames of a current chunk as well as to the first half of all previous chunks, which is different to other CSA methods \cite{TsunooKW21,WuWSY2020}. %
At the last encoder layer, to avoid redundancy due to overlap, only frames of the first half of each chunk are forwarded, except for the last chunk of an utterance for which all frames are maintained. The CSA latency is controlled by the chunk size.

\subsection{ASR Results}

\begin{table}[tb]
  \caption{ CERs and WERs for the HKUST and SWBD ASR tasks. }
  \label{tab:results_hkust_swbd}
  \centering
     \sisetup{table-format=2.1,round-mode=places,round-precision=1,table-number-alignment = center,detect-weight=true,detect-inline-weight=math}
  \resizebox{.97\linewidth}{!}
  {\setlength{\tabcolsep}{4pt}
  \begin{tabular}{ccccccccS}
  \toprule
  \multicolumn{3}{c}{Encoder} & \multicolumn{2}{c}{Decoder} & \multicolumn{2}{c}{HKUST [\%]} & \multicolumn{2}{c}{SWBD [\%]} \\
  Self-attention & {Delay} & {Type} & {Type} & {Delay} & {dev} & {test} & {callhm} & {swbd} \\
  \cmidrule(lr){1-3}\cmidrule(lr){4-5}\cmidrule(lr){6-7}\cmidrule(lr){8-9}
  Full-seq. & {Full} & Tr & {Offline} & {Full} & 20.7 & 21.3 & 17.3 & 8.6 \\
  Full-seq. & {Full} & Tr & {TA} & {Full} & 20.5 & 21.2 & 17.7 & 9.1 \\
  Full-seq. & {Full} & Tr & {TA} & {480~ms} & 20.3 & 21.0 & 17.7 & 8.8 \\
    RSA & {480~ms} & Tr & {Offline} & {Full} & 22.9 & 23.2 & 18.5 & 9.2 \\
  CSA  & {480~ms} & Tr & {Offline} & {Full} & 22.0 & 22.2 & 17.3 & 8.7 \\
  DCN w/o KD   & {480~ms} & Tr & {Offline} & {Full} & 22.4 & 22.5 & 18.2 & 8.9 \\
  DCN  & {480~ms} & Tr & {Offline} & {Full} & 22.1 & 22.3 & 17.6 & 8.9 \\
  [0.4ex]
\hdashline\noalign{\vskip 0.65ex}
  Full-seq. & {Full} & Co & {Offline} & {Full} & 19.1 & 20.1 & 14.8 & 7.1 \\ %
  Full-seq. & {Full} & Co & {TA} & {Full} & 18.8 & 19.9 & 14.9 & 7.3 \\ %
  RSA & {480~ms} & Co & {Offline} & {Full} & 21.2 & 22.0 & 18.3 & 8.5 \\ %
  DCN  & {480~ms} & Co & {Offline} & {Full} & 20.2 & 21.2 & 16.9 & 8.1 \\ %
[0.4ex]
\hline\noalign{\vskip 0.65ex}
  RSA   & {480~ms} & Tr & {TA} & {480~ms} & 23.1 & 23.4 & 19.6 & 10.2 \\
  CSA    & {480~ms} & Tr & {TA} & {480~ms} & 22.2 & 22.4 & 18.4 & 9.3  \\
  CSA    & {480~ms} & Tr & {TA} & {320~ms} & 22.3 & 22.4 & 18.4 & 9.5 \\
  CSA    & {640~ms} & Tr & {TA} & {320~ms} & 21.8 & 21.9 & 18.0 & 9.1 \\
  DCN & {480~ms} & Tr & {TA} & {480~ms} & 22.3 & 22.2 & 18.6 & 9.5  \\
  DCN & {480~ms} & Tr & {TA} & {320~ms} & 22.3 & 22.3 & 18.7 & 9.8 \\
  DCN & {640~ms} & Tr & {TA} & {320~ms} & 21.9 & 22.0 & 18.9 & 9.8 \\
[0.4ex]
\hdashline\noalign{\vskip 0.65ex}
  RSA & {480~ms} & Co & {TA} & {480~ms} & 21.7 & 22.1 & 18.8 & 9.1 \\ %
  DCN  & {480~ms} & Co & {TA} & {480~ms} & 20.6 & 21.3 & 17.8 & 8.2 \\ %
  DCN  & {640~ms} & Co & {TA} & {320~ms} & \bfseries 20.6 & \bfseries 21.1 & \bfseries 17.2 & \bfseries 8.1 \\ %
  
\bottomrule
  \end{tabular}}
\end{table}

Results of offline or partially streaming ASR models are shown for HKUST, SWBD, and LibriSpeech in the top half of Tables~\ref{tab:results_hkust_swbd} and \ref{tab:results_libri}, whereas fully streaming ASR results are shown in the bottom half. Transformer (Tr) and conformer (Co) results are separated by dashed lines. ASR results of the full-sequence based encoder models demonstrate that TA decoding \cite{MoritzHR19c,MoritzHLR20} achieves comparable results to offline joint CTC/attention decoding \cite{watanabe2018espnet}. For HKUST and LibriSpeech, even slightly improved character error rates (CERs) and word error rates (WERs) can be observed with the TA decoding algorithm. %
Note that fine-tuning a full-sequence based decoder to a TA decoder with limited look-ahead improves ASR results considerably compared to training from scratch. For example, for an RSA-based encoder with one frame look-ahead per layer (480 ms delay) and a TA decoder with 12 frames look-ahead (480 ms delay) the test-clean/test-other WERs of LibriSpeech are improved from 3.1\%/8.1\% \cite{MoritzHLR20} (trained from scratch) to 2.9\%/7.6\% (fine-tuning, cf.\ Table~\ref{tab:results_libri}).
DCN and CSA outperform RSA across all experiments often by a large margin. The RSA encoder with 480~ms delay corresponds to 1 frame look-ahead for each of the 12 encoder blocks, since the RSA delay grows linearly with the number of RSA layers and the frame rate amounts to 40~ms.
DCN and CSA both avoid a growing delay with the number of encoder blocks. Thus, for the 480~ms and 640~ms encoder delays, DCN and CSA process 12 and 16 look-ahead frames at each block.

The effect of using knowledge distillation (KD) for the DCN-based encoder is shown in Table~\ref{tab:results_hkust_swbd} by comparing the ``DCN w/o KD'' and ``DCN'' results for the ``offline'' decoder. KD improves HKUST results by up to 0.3\% and SWBD results by up to 0.6\%.
The best streaming E2E ASR results are highlighted in bold, which are obtained using the DCN-based conformer architecture with 640~ms (encoder) + 320~ms (decoder) delay. %

\vspace{-0.1cm}
\subsection{Decoding Delay}
\vspace{-0.1cm}

Tables~\ref{tab:results_hkust_swbd} and \ref{tab:results_libri} show ASR results and algorithmic delays of the encoder and decoder neural networks. However, a model may learn to delay an output until it has seen sufficient context or the decoding algorithm may also contain mechanisms to postpone the recognition of ASR labels. %
Therefore, in this section, we analyze the emission delay of the TA decoding algorithm, which is measured by computing the time difference between the word-level forced alignments provided by \cite{alignments_librispeech,montreal_forced_align} and the trigger positions of the TA decoding when recognizing a complete word. The emission delays are estimated across all four test sets of LibriSpeech for all words of sentences that are recognized correctly using $delay_w = t^\mathrm{pred}_w - t^\mathrm{grdt}_w$, where $w$ denotes a word ID, $t^\mathrm{pred}_w$ the predicted time position, and $t^\mathrm{grdt}_w$ the ground-truth time position of the word end.

\begin{table}[tb]
  \caption{ WERs [\%] for the LibriSpeech ASR task. * indicates usage of the Tr-LM for decoding. }
  \label{tab:results_libri}
  \centering
     \sisetup{table-format=2.1,round-mode=places,round-precision=1,table-number-alignment = center,detect-weight=true,detect-inline-weight=math}
  \resizebox{.97\linewidth}{!}
  {\setlength{\tabcolsep}{4pt}
  \begin{tabular}{ccccccccc}
  \toprule
  \multicolumn{3}{c}{Encoder} & \multicolumn{2}{c}{Decoder} & \multicolumn{2}{c}{dev} & \multicolumn{2}{c}{test} \\
  Self-attention & {Delay} & {Type} & {Type} & {Delay} & {clean} & {other} & {clean} & {other} \\
  \cmidrule(lr){1-3}\cmidrule(lr){4-5}\cmidrule(lr){6-7}\cmidrule(lr){8-9}
  Full-seq. & {Full} & Tr & {Offline} & {Full} & 2.4 & 5.9 & 2.6 & 5.9 \\
  Full-seq. & {Full} & Tr & {TA} & {480~ms} & 2.3 & 5.9 & 2.6 & 6.2 \\
  RSA & {480~ms} & Tr & {Offline} & {Full} & 2.5 & 6.7 & 2.8 & 7.1 \\
  CSA  & {480~ms} & Tr & {Offline} & {Full} & 2.4 & 6.6 & 2.6 & 6.7 \\
  DCN  & {480~ms} & Tr & {Offline} & {Full} & 2.5 & 6.7 & 2.6 & 6.8 \\
[0.4ex]
\hdashline\noalign{\vskip 0.65ex}
  Full-seq. & {Full} & Co & {Offline} & {Full} & 2.6 & 5.7 & 2.8 & 6.0 \\ %
  Full-seq. & {Full} & Co & {TA} & {Full} & 2.1 & 5.6 & 2.3 & 5.5 \\ %
  RSA & {480~ms} & Co & {Offline} & {Full} & 3.2 & 6.7 & 3.2 & 7.3 \\ %
  DCN  & {480~ms} & Co & {Offline} & {Full} & 2.9 & 6.5 & 3.0 & 6.7 \\ %
\hline\noalign{\vskip 0.65ex}
  RSA   & {0~ms} & Tr & {TA} & {480~ms} & 2.9 & 7.9 & 3.1 & 8.0 \\
  RSA   & {480~ms} & Tr & {TA} & {480~ms} & 2.7 & 7.3 & 2.9 & 7.6 \\
  CSA    & {480~ms} & Tr & {TA} & {480~ms} & 2.6 & 7.1 & 2.7 & 7.3  \\
  CSA    & {640~ms} & Tr & {TA} & {320~ms} & 2.6 & 6.9 & 2.8 & 7.3  \\
  DCN & {480~ms} & Tr & {TA} & {480~ms} & 2.7 & 7.1 & 2.8 & 7.4  \\
  DCN & {640~ms} & Tr & {TA} & {320~ms} & 2.6 & 6.9 & 2.8 & 7.3  \\
  [0.4ex]
\hdashline\noalign{\vskip 0.65ex}

  RSA & {480~ms} & Co & {TA} & {480~ms} & 2.5 & 6.7 & 2.7 & 7.1 \\ %
  DCN  & {640~ms} & Co & {TA} & {320~ms} & 2.4 & 6.6 & 2.6 & 6.7 \\ %
  {*\phantom{***}DCN\phantom{****}} & {640~ms} & Co & {TA} & {320~ms} & \bfseries 2.2 & \bfseries 6.4 & \bfseries 2.5 & \bfseries 6.3 \\ %

\bottomrule
  \end{tabular}}
   \vspace{-3mm}
\end{table}

\begin{figure}[t]
  \centering
  \centerline{\includegraphics[width=0.99\linewidth]{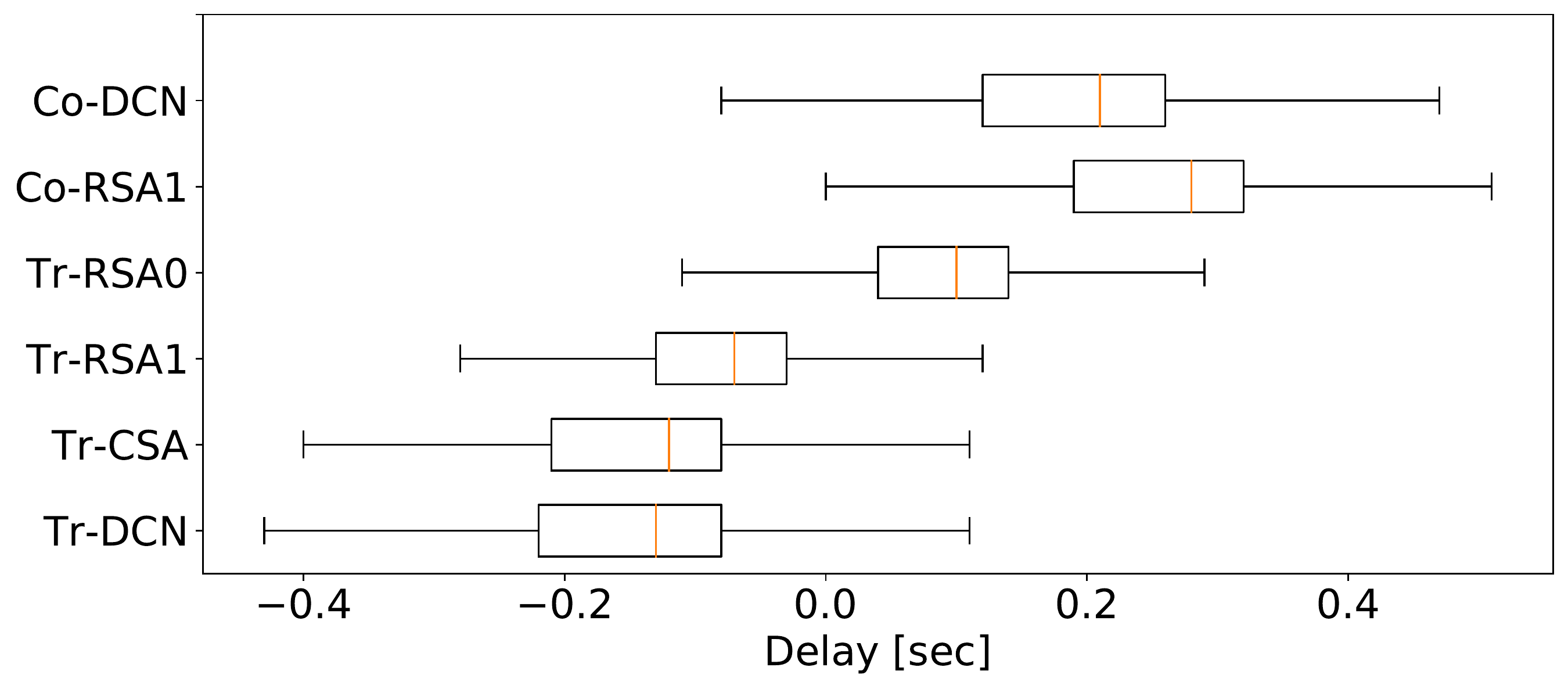}} %
  \caption{Box and whisker plot of word emission delays for TA-based ASR decoding using different encoder architectures. %
  All encoder and decoder models use the same look-ahead context of 480~ms except Tr-RSA0, where the encoder uses 0 look-ahead. } \vspace{.1cm}
\label{fig:dec_delay}
\end{figure}

Figure~\ref{fig:dec_delay} shows the decoding delays of different E2E ASR models, via a box and whisker plot summarizing the emission delays estimated from more than 100k words for each ASR setup.
Apparently, TA decoding, where timing is mainly defined by CTC, does not artificially %
delay ASR outputs, since recognized words are overall well aligned with the forced alignment results of a hybrid ASR system \cite{montreal_forced_align}.
Furthermore, by comparing ``Tr-DCN", ``Tr-CSA", and ``Tr-RSA1", which all use the same encoder and decoder delays of 480~ms, it can be noticed that DCN and CSA tend to produce less delay than RSA. We suppose the reason is that ASR models with less or no look-ahead tend to delay the emission of ASR labels, as can be seen from the RSA-based models with no look-ahead (``Tr-RSA0") and with 1 frame look-ahead per layer (``Tr-RSA1"), where the receptive field is not constant but linearly increases with the depth of the model. In addition, it is shown that conformer-based models generate larger ASR delays than transformer-based models. Our assumption is that the causal $\mathrm{Conv}$ module of the streaming conformer is the culprit due to a similar reason as discussed above regarding ``Tr-RSA0".

\section{Conclusions}

We presented the dual causal/non-causal (DCN) self-attention architecture for streaming ASR, which performs causal and non-causal self-attention simultaneously in a frame-synchronous manner. DCN self-attention uses causal frames to prevent self-attention from using information beyond the attention context and to avoid a growing receptive field and latency for multiple consecutive layers.
Combined with triggered attention (TA), the proposed streaming E2E ASR system demonstrates very strong results for various ASR tasks, e.g., achieving WERs of 2.5\%/6.3\% for the test-clean/test-other conditions of LibriSpeech with less than 1 second algorithmic delay. 
In addition, TA training by fine-tuning is shown to be effective, and the TA decoding latency is analyzed, demonstrating that ASR models with no or non-uniform look-ahead such as RSA are more prone to delaying the emission of ASR outputs.

\bibliographystyle{IEEEtran}

\bibliography{refs}

\end{document}